\documentclass[aps,prl,reprint,superscriptaddress]{revtex4-1}

\newcommand{\bra}[1]{\langle #1 |}
\newcommand{\ket}[1]{| #1 \rangle }

\newcommand{\bbraket}[3]{ \langle #1 | #2 | #3 \rangle }

\newcommand{\abs}[1]{\left \lvert #1 \right \rvert}

\DeclareRobustCommand{\quickfig}[4]{
\begin{figure}
\begin{centering}
\includegraphics[width=#1]{#2}
\par\end{centering}
\caption{#3}
\label{#4}
\end{figure}
}

\usepackage{amsmath}
\usepackage{amstext}
\usepackage{amssymb}
\usepackage{graphicx}
\usepackage{import}
\usepackage{appendix}

\usepackage{textcomp}

\usepackage[pdfpagemode=UseNone,pdfstartview=FitH,colorlinks=true,linkcolor=blue,citecolor=blue,urlcolor=blue]{hyperref}
\usepackage[all]{hypcap}

\begin{document}

\title{Measurement-Induced State Transitions in a Superconducting Qubit:\\Beyond the Rotating Wave Approximation}

\newcommand{\quantumai}{Google Inc., Santa Barbara, California 93117, USA}
\newcommand{\ucsb}{Department of Physics, University of California, Santa Barbara, California 93106-9530, USA}
\newcommand{\riversideee}{Department of Electrical and Computer Engineering, University of California, Riverside, California 92521, USA}
\newcommand{\riversidephysics}{Department of Physics, University of California, Riverside, California 92521, USA}

\author{Daniel Sank} \thanks{These authors contributed equally to this work}
\affiliation{\quantumai}

\author{Zijun Chen} \thanks{These authors contributed equally to this work}
\affiliation{\ucsb}

\author{Mostafa Khezri} \thanks{These authors contributed equally to this work}
\affiliation{\riversideee}
\affiliation{\riversidephysics}

\author{J. Kelly}
\affiliation{\quantumai}

\author{R. Barends}
\affiliation{\quantumai}

\author{B. Campbell}
\affiliation{\ucsb}

\author{Y. Chen}
\affiliation{\quantumai}

\author{B. Chiaro}
\affiliation{\ucsb}

\author{A. Dunsworth}
\affiliation{\ucsb}

\author{A. Fowler}
\affiliation{\quantumai}

\author{E. Jeffrey}
\affiliation{\quantumai}

\author{E. Lucero}
\affiliation{\quantumai}

\author{A. Megrant}
\affiliation{\quantumai}

\author{J. Mutus}
\affiliation{\quantumai}

\author{M. Neeley}
\affiliation{\quantumai}

\author{C. Neill}
\affiliation{\ucsb}

\author{P. J. J. O'Malley}
\affiliation{\ucsb}

\author{C. Quintana}
\affiliation{\ucsb}

\author{P. Roushan}
\affiliation{\quantumai}

\author{A. Vainsencher}
\affiliation{\quantumai}

\author{T. White}
\affiliation{\quantumai}

\author{J. Wenner}
\affiliation{\ucsb}

\author{Alexander N. Korotkov}
\affiliation{\riversideee}

\author{John M. Martinis}
\affiliation{\quantumai}
\affiliation{\ucsb}

\begin{abstract}
Many superconducting qubit systems use the dispersive interaction between the qubit and a coupled harmonic resonator to perform quantum state measurement.
Previous works have found that such measurements can induce state transitions in the qubit if the number of photons in the resonator is too high. 
We investigate these transitions and find that they can push the qubit out of the two-level subspace, and that they show resonant behavior as a function of photon number.
We develop a theory for these observations based on level crossings within the Jaynes-Cummings ladder, with transitions mediated by terms in the Hamiltonian that are typically ignored by the rotating wave approximation.
We find that the most important of these terms comes from an unexpected broken symmetry in the qubit potential.
We confirm the theory by measuring the photon occupation of the resonator when transitions occur while varying the detuning between the qubit and resonator.
\end{abstract}

\maketitle

The Jaynes-Cummings (JC) Hamiltonian \cite{Jaynes:1963, Tavis:1968} describes the interaction between a quantum two-level system and a harmonic oscillator, and is used to model a huge variety of physical systems.
For example, in superconducting qubits, it describes the interaction between the qubit and a resonator used to measure the qubit's state.
As predicted by the dispersive limit of the JC model, each qubit state induces a different frequency shift in the resonator, and the qubit state is inferred by measuring the resonator's response to a probe pulse \cite{Blais:cQED2004, Wallraff:2004, Koch:transmon2007}.
Dispersive measurement itself played a key role in recent experiments exploring the nature of quantum measurement \cite{Murch:trajectories2013, Weber:mapping2014, Hatridge:backaction2013}, and the high speed and accuracy of dispersive measurement has been critical in establishing superconducting qubits as a compelling technology for quantum computation \cite{Barends:gates2014,Chow:strand2014}.
Furthermore, repetitive error protection and characterization protocols \cite{Yan:flux_noise2012, Kelly:repetition_code2015, Corcoles:error_detection2015, Riste:bit_flips2015, Riste:charge_fluctuations2013, Fowler:surface_code2012} require that the qubit remain in a known state within the qubit subspace after the measurement is complete, a property guaranteed by the dispersive JC Hamiltonian.

However, several experiments with superconducting qubits have found that as the number of photons occupying the resonator $\bar{n}$ is increased past a certain point, the qubit suffers anomalous state transitions \cite{Reed:highPower2010, Johnson:heralding2012, Sank:thesis2014, Jeffrey:fast2014}.
It was long believed that these transitions could be explained by the breakdown of the dispersive approximation of the JC model as $\bar{n}$ exceeds a critical photon number $n_c$, but recent theory showed that the transitions are not predicted by the JC interaction even with very large $\bar{n}$ \cite{Khezri2016}.
Perhaps more puzzling, the transition probability is observed to be non-monotonic with increasing photon number.
As these transitions limit the speed and lower the fidelity of qubit measurement \cite{Jeffrey:fast2014, Johnson:heralding2012}, understanding and eliminating them is an important step in implementing high fidelity quantum algorithms, simulation, and error corrected computation.

In this Letter, we investigate the cause of anomalous qubit transitions in a superconducting qubit-resonator system.
We characterize the transitions by measuring the state of the qubit after driving the resonator with variable power, and find that the qubit jumps outside the two-level subspace.
Moreover, these transitions show a resonant behavior as a function of drive power.
By re-examining an important assumption of the JC Hamiltonian, namely the rotating wave approximation (RWA), we develop a theory based on level crossings with other states of the qubit-resonator system, and find that the theory matches experimental observations with no free parameters.

\begin{figure}
\begin{centering}
\includegraphics[width=\columnwidth]{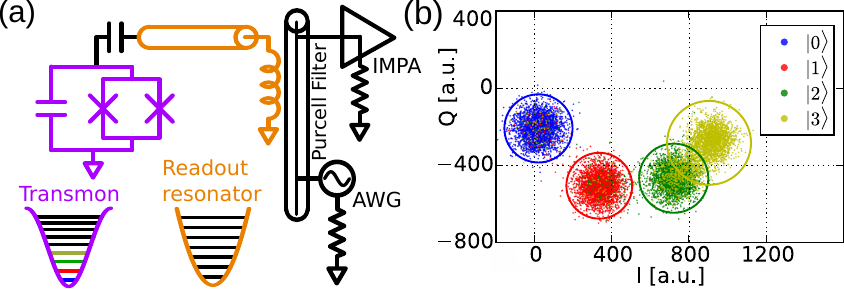}
\end{centering}
\caption{
Transmon-resonator system.
(a) Circuit and potential diagrams.
The transmon (violet) is capacitively coupled to the resonator (orange).
The resonator is inductively coupled to a bandpass Purcell filter with $Q \approx 30$ \cite{Jeffrey:fast2014}.
The resonator is driven by an arbitrary waveform generator connected to the filter, and the dispersed photons are measured by a low noise, impedance matched parametric amplifier \cite{Mutus:IMPA2014} also connected to the filter.
(b) In-phase and quadrature (IQ) components of the dispersed signal measured with the transmon prepared in the first four states, with each state forming an IQ ``cloud''.
The circles represent $3\sigma$ from fitting a Gaussian distribution to each cloud's projection onto lines connecting the clouds' centers.
}
\label{fig.system}
\end{figure}

Our experiment used a superconducting transmon qubit \cite{Koch:transmon2007, Barends:XMon2013} capacitively coupled to the fundamental mode of a quarter wave coplanar waveguide resonator with coupling strength $g/2\pi \approx 87 \, \text{MHz}$ \cite{g_frequency_dependence_note}, as illustrated in Fig.\,\ref{fig.system}\,(a).
The transmon's weakly anharmonic potential supports a ladder of energy levels, the bottom two of which are used as a qubit.
By biasing the transmon's dc SQUID with a magnetic flux, we can tune the transmon's $\ket{0} \rightarrow \ket{1}$ transition frequency $\omega_{10}$.
In the absence of bias flux, the transmon has its maximum frequency $\omega_{10}/2\pi = 5.4 \, \text{GHz}$, and the anharmonicity is $\eta / 2\pi \equiv (\omega_{21} - \omega_{10})/2\pi = -221 \, \text{MHz}$.
The fundamental mode of the resonator is a quantum harmonic oscillator with frequency $\omega_r / 2\pi \approx 6.78 \, \text{GHz}$ and is coupled with an energy decay rate of $\kappa \approx 1 / (37\,\text{ns})$ through a bandpass Purcell filter  \cite{Reed:filter2010, Jeffrey:fast2014} to a $50\,\Omega$ output line and amplifiers.

Each transmon level $\ket{i}$ induces a different frequency shift on the resonator, yielding a set of distinct resonator frequencies $\omega_{r,\ket{i}}$.
To measure the transmon state, we drive the system through the Purcell filter at a frequency between $\omega_{r,\ket{1}}$ and $\omega_{r,\ket{2}}$ \cite{readout_frequency_note}, populating the resonator with photons that leak out from the resonator, through the filter, and into the amplifier circuit.
The amplitude and phase of the outgoing photons are shifted (dispersed) in a way that depends on the resonator frequency, and thus the transmon state.
We digitize this signal and extract the amplitude and phase as a point in the IQ plane.
In Fig.\,\ref{fig.system}\,(b), we plot the IQ response of the resonator with the transmon prepared in various states, which acts as our calibration for distinguishing the state of the transmon in subsequent measurements.
When the resonator-transmon detuning $\abs{\Delta} \equiv \abs{\omega_{10} - \omega_r}$ is not more than $1.4\,\text{GHz}$, the resulting IQ points resolve up to the first four transmon states, while at larger $\abs{\Delta}$ (relevant to most of our data) we can only resolve the first three states due to the smaller dispersive shift.

\begin{figure}
\begin{centering}
\includegraphics[width=\columnwidth]{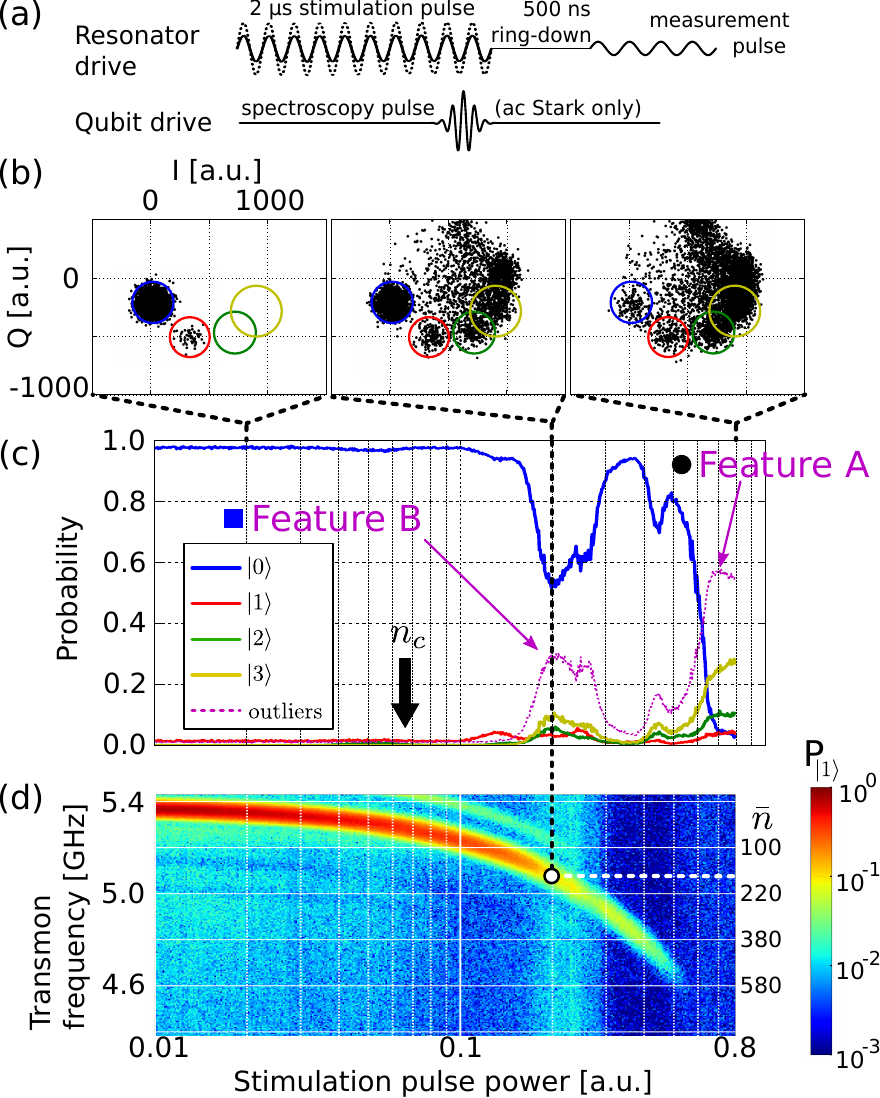}
\end{centering}
\caption{
(a) Control sequence for probing the effect of resonator photons on the transmon.
The spectroscopy pulse is used only in the ac Stark measurement.
(b) IQ data for drive powers 0.02, 0.2, and 0.8 (arbitrary units), with $\omega_{10} = 5.38\,\text{GHz}$.
The circles represent $3\sigma$ for the four resolvable transmon states as calibrated in Fig.\,\ref{fig.system}b.
At high power, the transmon is clearly driven to states higher than $\ket{3}$.
(c) Transmon state probabilities versus stimulation power.
In addition to the four calibrated transmon states, we show the probability that the measurement was $>3\sigma$ from any of the resolved states, labeled ``outliers''.
Note the two large resonance-like peaks labeled $A$ and $B$.
(d) Stark shifted transmon frequency $\omega_{10}$ versus stimulation pulse power.
We convert the shifted $\omega_{10}$ to $\bar{n}$ using a numerical theory (right vertical axis) \cite{supplement}.
}
\label{fig.iq_data}
\end{figure}

To investigate the effect of resonator photons on the transmon state, we use the pulse sequence illustrated in Fig.\,\ref{fig.iq_data}\,(a).
The transmon is initialized to $\ket{0}$ by idling for several times its energy decay lifetime.
We first drive the resonator with a $2\,\mu\text{s}$ long, variable power pulse.
This ``stimulation pulse'' injects a number of photons into the resonator that, when large enough, induces transitions in the transmon state.
We then wait $500\,\text{ns}$ (13 decay time constants) for the resonator to ring down \cite{qubit_t1_note}.
Finally, we drive the resonator again with a fixed low power pulse to measure the transmon without inducing further transitions, and record the IQ response of the resonator.
Based on the calibration shown in Fig\,\ref{fig.system}d, we identify each IQ point as one of the transmon states, or if the point is more than three standard deviations from any of the calibrated distributions, we label it as an ``outlier''.

The results are striking in two ways.
First, as the stimulation pulse power is raised, the transmon jumps from $\ket{0}$ not only to $\ket{1}$ but also to $\ket{2}$, $\ket{3}$ and even higher states, as shown in Fig.\,\ref{fig.iq_data}\,(b).
Although we can resolve only up to $\ket{3}$, the characteristic arc of the IQ points with increasing state index appears to continue to what we estimate to be $\ket{5}$ or higher.
Second, the probability of transitions is highly non-monotonic with power, as was previously seen in Refs. \cite{Sank:thesis2014, Johnson:heralding2012}.
In particular, the shapes of the features in probability versus power resemble resonance peaks, with large peaks in the outlier probability at drive powers 0.7 (feature $A$) and 0.2 (feature $B$), a small peak in $\ket{1}$ near 0.15, another small peak in $\ket{2}$ near 0.05, and various other peaks at other powers.
The peaked structure rules out any process that would have monotonically increasing transitions with increasing drive power, such as chip heating or dressed dephasing \cite{Boissonneault:DressedDephasing2008,Slichter2012}, as the dominant mechanism.

In order to connect our results to theoretical models, we next convert stimulation pulse power to photon number $\bar{n}$.
We cannot measure $\bar{n}$ directly, but resonator photons cause the qubit frequency to shift downward in what is called the ac Stark effect \cite{Schuster:acStarkDephasing2005}.
We map drive power to $\bar{n}$ by measuring the ac Stark shifted qubit frequency for each resonator drive power and converting that frequency to $\bar{n}$ using a numerical model based on separately measured parameters $g$ and $\Delta$ \cite{supplement}.
To measure the ac Stark shift, we repeat the previous experiment with the addition of a spectroscopic microwave pulse on the transmon after the driven resonator has reached the steady state.
For each drive power we vary the frequency of the transmon pulse; the $\ket{1}$ probability is maximized when the pulse is on resonance with the shifted transmon frequency.

We show the results of the ac Stark shift measurement with the computed photon numbers in Fig.\,\ref{fig.iq_data}\,(d) for the same drive powers as in Fig.\,\ref{fig.iq_data}\,(c).
Note that feature $B$ (black dashed line) occurs at $170 \lesssim \bar{n} \lesssim 250$, which is, interestingly, considerably larger than the critical photon number $n_c \equiv (\Delta / g)^2/4 \approx 60$ introduced in Ref.\,\cite{Blais:cQED2004}.

The peaks in Fig.\,\ref{fig.iq_data}\,(c) are thus seen to indicate particular values of $\bar{n}$ at which the qubit-resonator system is especially susceptible to transitions.
The association of $\bar{n}$ with qubit frequency shift further suggests that the peaks are due to some form of frequency resonance.
With the observation of resonant transitions to higher transmon levels, we now consider the Hamiltonian of the transmon-resonator system and look for terms, possibly neglected in the dispersive or rotating wave approximations, which explain these observations.

\begin{figure}
\begin{centering}
\includegraphics[width=\columnwidth]{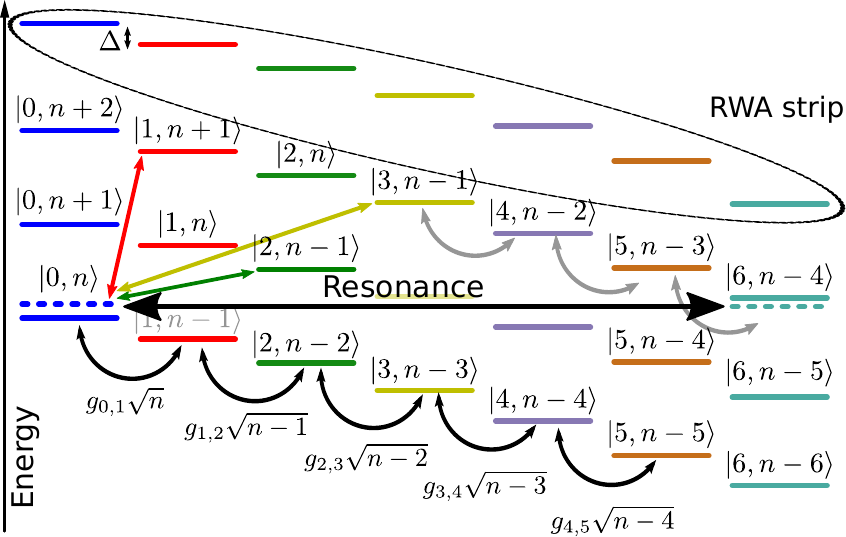}
\end{centering}
\caption{
JC ladder for large values of $n$.
Bare states are shown as solid lines and two of the eigenstates are shown as dashed lines.
Dark curved arrows indicate coupling within an RWA strip with corresponding RWA coupling strengths shown below.
The ladder has an energy resonance between $\ket{0,n}$ and $\ket{6,n-4}$ (long black arrow).
Non-RWA couplings (short straight arrows) allow for interstrip transitions.
The couplings to $\ket{1,n+1}$ (red) and $\ket{3,n-1}$ (yellow), along with those within the RWA strip, mediate the transition between the resonant levels.
The coupling to $\ket{2,n-1}$ (green), which mediates additional resonant transitions, requires a Hamiltonian term coupling transmon states of equal parity; this is forbidden if the transmon potential is symmetric.
Note the energy spacing between states $\ket{k,n}$ and $\ket{k+1, n-1}$ is $\Delta$ as indicated in the top left.
}
\label{fig.ladder}
\end{figure}

We start with the bare Hamiltonian
\begin{equation}
H_\text{b} = \sum_k E_k \ket{k}\bra{k} + \hbar \omega_r a^\dagger a \, ,
\end{equation}
where $E_k$ is the energy of transmon level $k$ and $\omega_r$ is the frequency of the resonator.
This Hamiltonian produces the JC ladder as shown by the solid lines in Fig.\,\ref{fig.ladder}.

Adding the interaction term $H_\text{I}$ due to the capacitive coupling gives
\begin{equation}
H_\text{I} = \sum_{k,k',n} \hbar g_{k,k'} \sqrt{n} \, \ket{k', n-1}\bra{k, n} + \text{H.c.} \, ,
\end{equation}
where the states are labeled $\ket{\text{qubit}, \text{resonator}}$, $g_{k,k'} = g \, \bbraket{k}{Q}{k'} / \bbraket{0}{Q}{1}$, and $\bbraket{k}{Q}{k'}$ are the transmon charge matrix elements.
This interaction imparts an $n$-dependent shift on the bare levels producing eigenstates, two of which are shown as dashed lines in Fig.\,\ref{fig.ladder}.
As indicated by the long horizontal arrow, at certain $n$ the ladder contains resonances between states where the qubit goes from $\ket{0}$ to higher levels such as $\ket{6}$.
This critical observation could explain both the resonance structure and the transitions to higher transmon levels observed in the data.
However, it remains to see how $H_\text{I}$ couples the resonant levels.

The full interaction $H_\text{I}$ is typically simplified by the RWA to contain only those terms that preserve excitation number,
\begin{equation}
H_\text{RWA} \equiv \sum_{k,n} \hbar g_{k, k+1} \sqrt{n} \, \ket{k+1, n-1}\bra{k,n} + \text{H.c.} \, .
\end{equation}
These terms (curved arrows in Fig.\,\ref{fig.ladder}) divide the JC ladder into excitation preserving subspaces which we call ``RWA strips''.
Under $H_\text{RWA}$, the system moves only \emph{within} an RWA strip; taking the system out of the dispersive limit with $n \gg n_c$ \emph{only} results in a reduction of the resonator dispersive shift \cite{Khezri2016, supplement}.
Therefore, $H_\text{RWA}$ does not allow transitions between resonant levels.

The critical part of the Hamiltonian is $H_\text{non-RWA}$, containing terms in $H_\text{I}$ that do not conserve excitation number \cite{supplement}.
These terms can be as large as the RWA terms, but are usually neglected on the grounds that they are more off resonant than the RWA terms (in our system the RWA terms are $\sim 1\,\text{GHz}$ off resonance, while the non-RWA terms are $\sim 13\,\text{GHz}$ off resonance).
However, keeping these terms reveals the essential reason for the unwanted state transitions.
The non-RWA terms couple next-nearest neighboring RWA strips (i.e., those differing by 2 in total excitation number) together, as shown in Fig.\,\ref{fig.ladder}.
Combined with the intrastrip coupling provided by $H_\text{RWA}$, the non-RWA coupling allows multistep (i.e., higher order) processes to connect the resonant levels.
For example, $H_\text{non-RWA}$ carries the system from $\ket{0, n}$ to $\ket{1, n+1}$ in another RWA strip, and then $H_\text{RWA}$ carries the system within the strip to $\ket{6, n-4}$.
Note that although the full process conserves energy, the individual steps do not.

\begin{figure}
\begin{centering}
\includegraphics[width=\columnwidth]{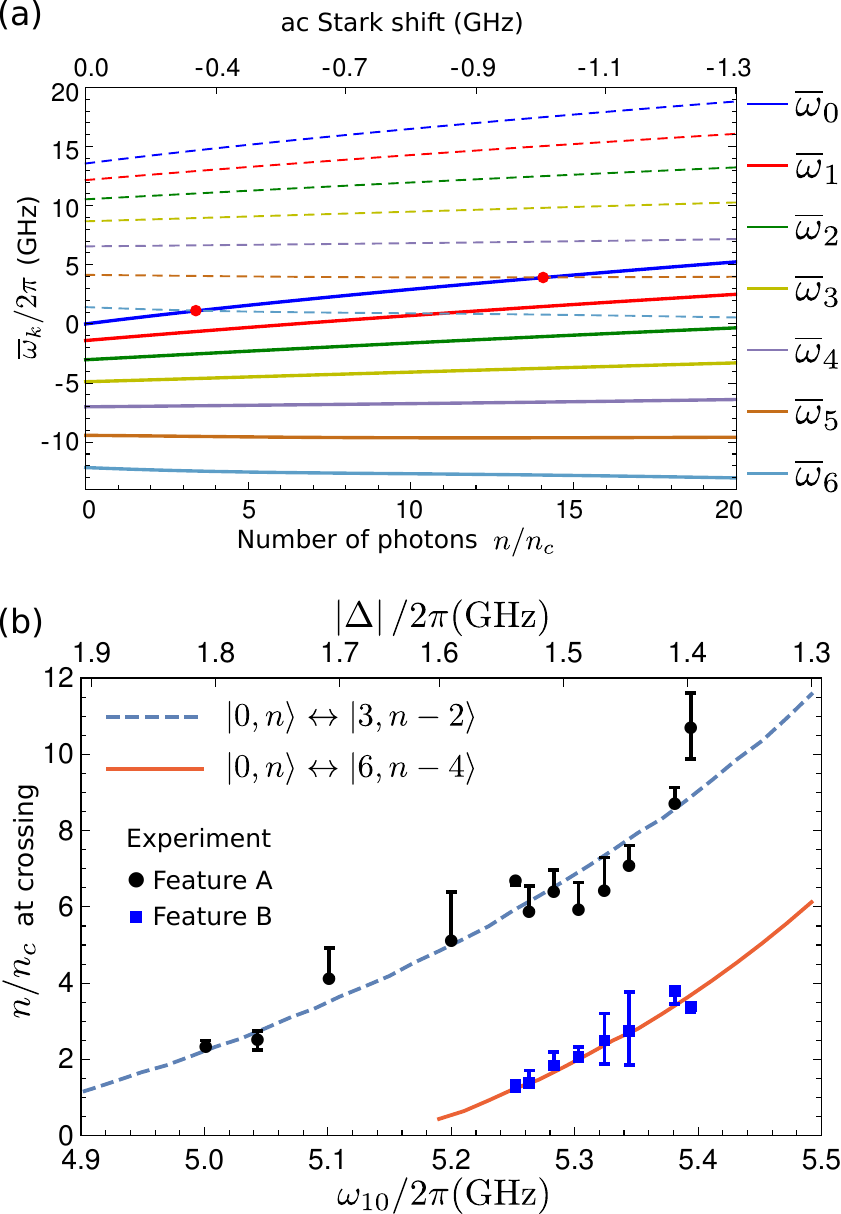}
\end{centering}
\caption{
(a) Fan diagram of the energy levels within an RWA strip.
Solid: Frequencies $\overline{\omega}_k (n) \equiv E_{\overline{\ket{k,n-k}}} / \hbar - n \omega_r$ versus photon number $n$ for $\abs{\Delta}=1.4 \, \text{GHz}$.
As $n$ increases, the levels repel more strongly and fan out.
Dashed: Same frequencies shifted by $2 \omega_r$, which represent the next-nearest neighboring RWA strip.
The red dots show energy resonances with the qubit state $|0\rangle$ occurring at specific values of $n$.
The left dot corresponds to the resonance shown in Fig\,\ref{fig.ladder}.
(b) Photon number at level crossing versus $\omega_{10}$, compared between experiment and theory.
Black circles and blue squares show experimental features A and B from Fig.\,\ref{fig.iq_data} respectively, and the error bars represent the apparent widths of the features.
Solid red line is the theory prediction for level crossing between eigenlevels of $\ket{0,n}$ and $\ket{6,n-4}$.
Dashed blue line is the theory prediction for an asymmetric transmon that breaks the selection rule by at least $1\%$, yielding level crossings between eigenlevels of $\ket{0,n}$ and $\ket{3,n-2}$.
}
\label{fig.theory}
\end{figure}

To find the condition under which the resonances occur, we numerically compute the frequencies $\overline{\omega}_k (n) \equiv E_{\overline{\ket{k,n-k}}}/\hbar - n\omega_r$ (the overline indicates eigenstate) of the levels within each RWA strip, as functions of $n$.
As $n$ increases, energy levels within each strip repel each other more strongly and fan out, as illustrated by the solid lines in the ``fan diagram'' in Fig.\,\ref{fig.theory}\,(a).
By superimposing fan diagrams of two next-nearest neighboring RWA strips, as shown by the dashed lines, we see that they have multiple intersections, meaning that the JC ladder contains multiple resonances.
For example, the left red dot in Fig.\,\ref{fig.theory}\,(a) shows that the transmon-resonator state $\ket{0, n}$ can be brought on resonance with $\ket{6,n-4}$.
The presence of crossings with higher transmon states agrees with the experimental observation of transitions to states higher than $\ket{3}$.

Next, we compute the $n$ at which various intersections occur as a function of the qubit-resonator detuning $\Delta$, yielding the lines in Fig.\,\ref{fig.theory}\,(b).
As $\abs{\Delta}$ increases, the spacing between levels within an RWA strip also increases, see Fig.\,\ref{fig.ladder}.
However, the spacing between strips is fixed at $\omega_r$, so with increased $\abs{\Delta}$ fewer photons are required to bring $\ket{0, n}$ on resonance with states in higher strips and so the transitions occur at lower $\bar{n}$.
Note that while we use $n$ in the theory, the experiment drives the resonator into a coherent state with mean photon number $\bar{n}$ and fluctuations $\sqrt{\bar{n}} < 0.1 \, \bar{n}$.
Also, although the $n$ at which the energy resonance occurs is not related to $n_c$, the effective couplings between resonant levels are large enough to yield the experimental features only when $n \gtrsim n_c$.

To confirm the theoretical prediction, we repeat the experiment shown in Fig.\,\ref{fig.iq_data} for several values of $\omega_{10}$ by biasing the transmon's SQUID with magnetic flux.
At each $\omega_{10}$, we find the values of $\bar{n}$ of features $A$ and $B$, as shown in Fig.\,\ref{fig.iq_data}\,(d)), and plot these points in Fig.\,\ref{fig.theory}\,(b).
The experimental points for feature $A$ (black circles) and feature $B$ (blue squares) are well fit by numerically computed curves for the transitions from $\ket{0, n}$ to $\ket{6,n-4}$ and $\ket{3, n-2}$, respectively.
Note that the theory lines are calculated using only the measured $\omega_r$, $\omega_{10}$, and $g$, with no free parameters fitted to the data.

However, the transition from $\ket{0, n}$ to $\ket{3, n-2}$ is actually unexpected.
If the transmon potential is symmetric, as is usually assumed \cite{Koch:transmon2007}, then $g_{i,j}$ is only nonzero when $j-i$ is odd.
Therefore, $H_\text{I}$ should only couple RWA strips where the difference in total excitation number is even, so the transition to $\ket{3,n-2}$ should be forbidden.
Nevertheless, the theory line for the $\ket{3, n-2}$ transition fits the data well, indicating a possible asymmetry in the transmon potential.
We confirmed this asymmetry by observing $\ket{0} \rightarrow \ket{2}$ Rabi oscillations when driving the transmon at $\omega_{01} + \omega_{12}$ \cite{supplement}.
Through comparison with Rabi oscillations on the $\ket{0} \rightarrow \ket{1}$ transition, we experimentally estimate $\abs{\bbraket{0}{Q}{2}/\bbraket{0}{Q}{1}} \approx 10^{-2}$ \cite{supplement}.
This matrix element is large enough to explain the transitions to $\ket{3, n-2}$, and so the level crossing theory appears to correctly predict both of the largest resonance features observed in the data.

We note that any spurious TLS coupled to transmon-resonator system can also participate in level crossings, and can lead to similar features (possibly the small peaks in Fig.\,\ref{fig.iq_data}\,(c)), even at lower photon numbers \cite{supplement}.

In conclusion, we find that strong dispersive measurement of a transmon induces transitions to states above $\ket{3}$.
These transitions occur at specific values of the photon occupation in the measurement resonator, and are caused by energy resonances within the qubit-resonator system.
Coupling between the resonant levels is mediated by Hamiltonian terms usually dropped in the rotating wave approximation, and the most important such term involves an unexpected broken symmetry in the transmon potential.
An interesting consequence of these results is that a system with smaller $\abs{\Delta}$ should allow larger photon numbers before resonant transitions occur.
This observation could be critical to improving measurement accuracy in dispersively measured systems, and may explain the large photon numbers used in Ref.\,\cite{Bultink:depletion2016}.
This work suggests several further avenues of research: characterizing level crossings with the qubit initialized in $\ket{1}$, determining the mechanism for the transmon's broken symmetry, clarifying the role of TLSs in non-RWA transitions, and understanding the $n$-dependent rates of the non-RWA transitions.

\begin{acknowledgments}
The authors thank C. Bultink for an enlightening discussion.
This work was supported by Google.
Z. C. and C. Q. acknowledge support from the National Science Foundation Graduate Research Fellowship under Grant No. DGE 1144085.
M. K. and A. N. K. acknowledge support from ARO Grants No. W911NF-15-1-0496 and No. W911NF-11-1-0268.
Devices were made at the UC Santa Barbara Nanofabrication Facility, a part of the NSF-funded National Nanotechnology Infrastructure Network, and at the Nanostructures Cleanroom Facility.
\end{acknowledgments}

%

\clearpage 
\onecolumngrid 
\vspace{\columnsep}
\begin{center}
\textbf{\large Supplementary Information for Measurement-Induced State Transitions in a Superconducting Qubit: Beyond the Rotating Wave Approximation}
\end{center}
\vspace{\columnsep}
\twocolumngrid

\setcounter{equation}{0}
\setcounter{figure}{0}
\setcounter{table}{0}
\setcounter{page}{1}

\renewcommand{\theequation}{S\arabic{equation}}
\renewcommand{\thefigure}{S\arabic{figure}}
\renewcommand{\bibnumfmt}[1]{[S#1]}
\renewcommand{\citenumfont}[1]{S#1}

\renewcommand{\theHtable}{Supplement.\thetable}
\renewcommand{\theHfigure}{Supplement.\thefigure}

\section{Hamiltonian}

The Hamiltonian of the coupled qubit-resonator system can be written as
\begin{equation}\label{eq:H}
H = H_\text{b} + H_\text{I}
\end{equation}
where $H_\text{b}$ is the ``bare'' Hamiltonian of the qubit and resonator, while $H_\text{I}$ describes their capacitive coupling.
With the ket convention $\ket{\text{qubit, resonator}}$, the bare Hamiltonian has the form
\begin{equation}\label{eq:H_b}
H_\text{b} = \sum_{k,n} \left( E_k  + n \hbar \omega_r \right) \ket{k,n}\bra{k,n}
\end{equation}
where $\omega_r$ is the (bare) resonator frequency, and $E_k$ is the transmon energy of level $k$, calculated numerically using Mathieu characteristic functions \cite{Koch2007}.
The transmon transition frequencies are $\omega_{kl} \equiv (E_k - E_l) / \hbar$ and its anharmonicity is $\eta \equiv \omega_{21} - \omega_{10}$.
This bare Hamiltonian produces the Jaynes-Cummings (JC) ladder of energy levels, shown in Fig.\,3 in the main text.

The interaction Hamiltonian $H_\text{I}$, given by Eq.\ (2) in the main text, is due to charge-charge coupling between the resonator and transmon.
It can be divided into two parts,
\begin{equation}
H_\text{I} = H_\text{RWA} + H_\text{non-RWA} \, ,
\end{equation}
where $H_\text{RWA}$ contains only terms conserving total excitation number, while $H_\text{non-RWA}$ contains the rest of the terms.
$H_\text{RWA}$ has the form
\begin{equation}\label{eq:H_RWA}
H_\text{RWA} = \sum_{k,n} \hbar g_{k,k+1} \sqrt{n} \, \ket{k+1,n-1} \bra{k,n} + \text{H.c.}  \, ,
\end{equation}
where $g_{k,k'} \equiv g \bbraket{k}{Q}{k'}/\bbraket{0}{Q}{1}$ are the normalized matrix elements of the transmon charge operator $Q$.
These matrix elements are calculated numerically using Mathieu functions.
In the case $k' = k + 1$, the matrix elements are approximately (for not very large values of $k$)
\begin{equation}
g_{k,k+1} \approx g \sqrt{k+1} \left( 1 + \frac{\eta}{2 \omega_{10}} \, k \right) \, .
\end{equation}
By diagonalizing $H_\text{b} + H_\text{RWA}$, we find the eigenstates $\overline{\ket{k,n}}$ and eigenenergies $E_{\overline{\ket{k,n}}}$, which we use to numerically compute the frequencies $\overline{\omega}_k (n) = E_{\overline{\ket{k,n-k}}} / \hbar - n \omega_r$ within each RWA strip (see the fan energy diagram in Fig.\,4\,(a) in the main text).

From $H_\text{b} + H_\text{RWA}$ we also numerically compute the photon number dependent ac Stark shift $\delta \omega_{10} \equiv \left( E_{\overline{\ket{1,n}}} - E_{\overline{\ket{0,n}}} \right) / \hbar -\omega_{10}$, as illustrated in Fig.\,\ref{fig.stark}.
This map between resonator photon number and transmon ac Stark shift, which provides the calibration between drive power and photon number discussed in the main text, was the critical link between theory and experiment.
Notice that Eq.\ \eqref{eq:H_RWA} goes beyond the usual dispersive approximation \cite{Blais2004}.
In particular, the numerically computed curve deviates noticeably from the usual linear relation $\delta \omega_{10} = - 2 \abs{\chi} n$.

\quickfig{0.9\columnwidth}
{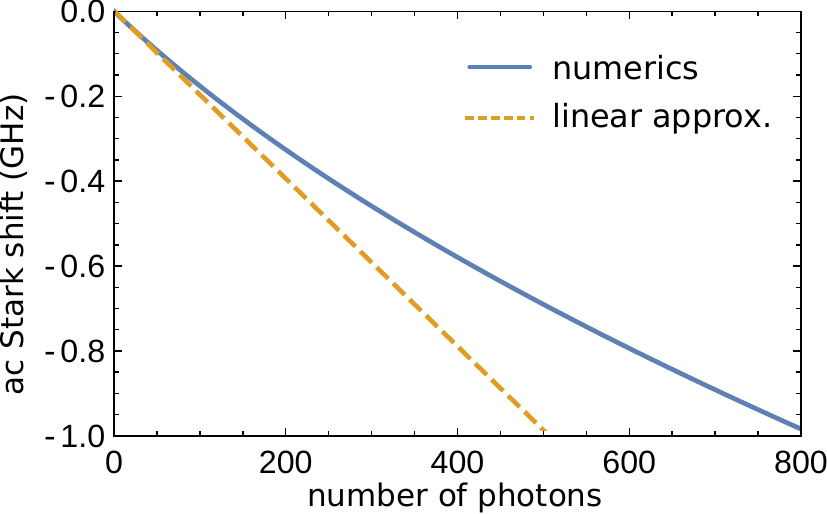}
{
Ac Stark shift of the transmon frequency as a function of the number of resonator photons $n$, for parameters of Fig.\,2 in the main text ($n_c \approx 60$), using $H_\text{RWA}$.
The solid line shows the value computed numerically, and the dashed line shows the conventional linear approximation $\delta \omega_{10} = -2 \abs{\chi} n$.
As $n$ becomes large, the relation between ac Stark shift and photon number becomes somewhat nonlinear.
}
{fig.stark}

The rest of the charge-charge interaction terms do not preserve excitation number, and are called here ``non-RWA'' terms.
Although some of these terms are as large as RWA terms, they are usually neglected since they are more off-resonant than RWA terms.
However, these terms connect RWA strips and therefore enable resonant transitions in the JC ladder, as explained in the main text.
In general, there are many types of non-RWA terms, which differ in coupling strength and in how close they are to resonantly connecting two JC ladder levels.
We only consider terms involving $g_{k,k+1}$ and $g_{k,k+3}$, as they are the largest and least off-resonant,
\begin{align}\label{eq:H_nonRWA}
H_\text{non-RWA}^{(1)}
= &\sum_{k,n}
\hbar \, g_{k,k+1}  \sqrt{n+1} \, \ket{k+1,n+1} \bra{k,n} + \text{H.c.}  \nonumber \\
+  &\sum_{k,n}  \hbar \, g_{k,k+3}  \sqrt{n} \, \ket{k+3,n-1} \bra{k,n} + \text{H.c.}  \, .
\end{align}
The couplings $g_{k,k+3}$ are calculated numerically; they are much smaller than $g_{k,k+1}$, as seen from the approximate formula
\begin{equation}
g_{k,k+3} \approx g \sqrt{(k+1)(k+2)(k+3)} \, \frac{-\eta}{4 \omega_{10}} \, .
\end{equation}
In spite of being relatively small, these couplings are numerically more important in our problem than couplings $g_{k,k+1}$. 
We note that $H_\text{non-RWA}$ induces slight changes in the eigenenergies $E_{\overline{\ket{k,n}}}$, but the effect is small enough that we neglect it.

Equation \eqref{eq:H_nonRWA} does not have any terms of the form $g_{k,k+2}$, and therefore only connects RWA strips differing in total excitation number by 2, which we call ``next-nearest neighbors'' (see Fig.\,3 in the main text).
The absence of $g_{k,k+2}$ terms is due to the symmetry of the transmon potential (in the phase basis).
However, the real system violates this selection rule (see Fig.\,\ref{fig.g_eff}\,(b) discussed later and also the discussion in the main text).
Accounting for the broken symmetry adds terms to $H_\text{non-RWA}$,
\begin{equation}\label{eq:H_nonRWA-2}
H_\text{non-RWA}^{(2)} = \sum_{k,n} \hbar \, g_{k,k+2} \, \sqrt{n}\, \ket{k+2,n-1}\bra{k,n}+ \text{H.c.} \, .
\end{equation}
The non-RWA terms of Eq.\ \eqref{eq:H_nonRWA-2} connect RWA strips differing in total excitation number by 1, which we call ``nearest neighbors'' (see Fig.\,3 in the main text), leading to additional resonance processes, such as $\overline{\ket{0,n}} \rightarrow \overline{\ket{3,n-2}}$.

\section{Effective Coupling}

When a resonance occurs between the initial state $\overline{\ket{0, n}}$ and, e.g., $\overline{\ket{6, n-4}}$, the system can make a resonant transition.
In the perturbative language, in making this transition the system goes through several intermediate off-resonant states (see Fig.\,3 in the main text); many different paths are available (i.e. different virtual processes).
As an example, one path is $\ket{0,n} \rightarrow \ket{1,n-1} \rightarrow \ket{4,n-2} \rightarrow \ket{5,n-3} \rightarrow \ket{6,n-4}$, which involves the matrix element $g_{1,4}$.
The condition of resonance is necessary but not sufficient to give these processes a measurably large probability; the process must also have large enough effective coupling between initial and final states.
We define the effective coherent coupling $g_\text{eff}^\text{coh}$ as
\begin{equation}\label{eq:g-eff-c}
g_\text{eff}^\text{coh}
= \overline{\bra{k_f, n_f}} H_\text{non-RWA} \overline{\ket{k_i, n_i}} \, ,
\end{equation}
where $\overline{\ket{k_i, n_i}}$ and $\overline{\ket{k_f, n_f}}$ are the initial and final eigenstates, respectively.
To find $g_\text{eff}^\text{coh}$, we expand the (RWA) eigenstates in the bare state basis, 
\begin{equation}
\overline{\ket{k,n}} = \sum_{l=0}^{k_{\text{max}}} c_l^{(k,n)} \ket{l, n+k-l},
\end{equation}
where $k_{\text{max}}\simeq 9$ is the highest transmon level taken into account.
This expansion is then substituted into Eq.\ (\ref{eq:g-eff-c}).
In particular, for the transition $\overline{\ket{0,n}}\rightarrow \overline{\ket{k,n-k+2}}$ (to the next-nearest neighboring RWA strip) the effective coupling is
\begin{align}\label{eq:g-eff-full}
g_\text{eff}^\text{coh} =
& \sum \nolimits_l c_l^{(0,n)} \hbar g_{l,l+1} \sqrt{n-l+1} \, \left[ c_{l+1}^{(k,n-k+2)} \right]^*  \nonumber \\
+ & \sum \nolimits_l c_l^{(0,n)} \hbar g_{l,l+3} \sqrt{n-l} \, \left[ c_{l+3}^{(k,n-k+2)} \right] ^* \, .
\end{align}
Each term in Eq.\ (\ref{eq:g-eff-full}) corresponds to a particular path in the picture of virtual processes.
The paths in the first line are $\ket{0,n} \rightarrow \ket{l,n-l}\rightarrow \ket{l+1,n-l+1}\rightarrow \ket{k,n-k+2}$, where the first and last arrows describe subpaths within the RWA strips.
Similarly, the terms in the second line correspond to paths $\ket{0,n} \rightarrow \ket{l,n-l} \rightarrow \ket{l+3,n-l-1} \rightarrow \ket{k,n-k+2}$.

The solid red line in Fig.\,\ref{fig.g_eff}\,(a) shows $g_\text{eff}^\text{coh}$ for the  $\overline{\ket{0,n}} \rightarrow \overline{\ket{6,n-4}}$ transition (so that $n$ corresponds to the resonance condition $E_{\overline{\ket{0,n}}} \approx E_{\overline{\ket{6,n-4}}}$), calculated using Eq.\ \eqref{eq:g-eff-c} or, equivalently, Eq.\ \eqref{eq:g-eff-full}.
Note that the terms in Eq.\ \eqref{eq:g-eff-full} are large at $n > n_c$ because $g_{l,l+1} \sqrt{n} \approx \abs{\Delta} \sqrt{l+1} \sqrt{n / 4 n_c}$ (typically a few GHz) and the amplitudes $c_l$ are significant for several states within the RWA strip.
Nevertheless, the result for $g_\text{eff}^\text{coh}$ shown by the solid red line in Fig.\,\ref{fig.g_eff}\,(a) is  smaller than even one such term.
The reason is an almost perfect cancellation of the terms in Eq.\ \eqref{eq:g-eff-full}, which happens because while the coefficients $c_l^{(k,n-k+2)}$ alternate in sign with changing $l$ for $l < k$, the coefficients $c_{l}^{(0,n)}$ are all positive \footnote{This follows from sequential perturbation theory}.
Therefore, the terms in Eq.\ \eqref{eq:g-eff-full} have alternating signs and efficiently cancel each other.

\quickfig{\columnwidth}
{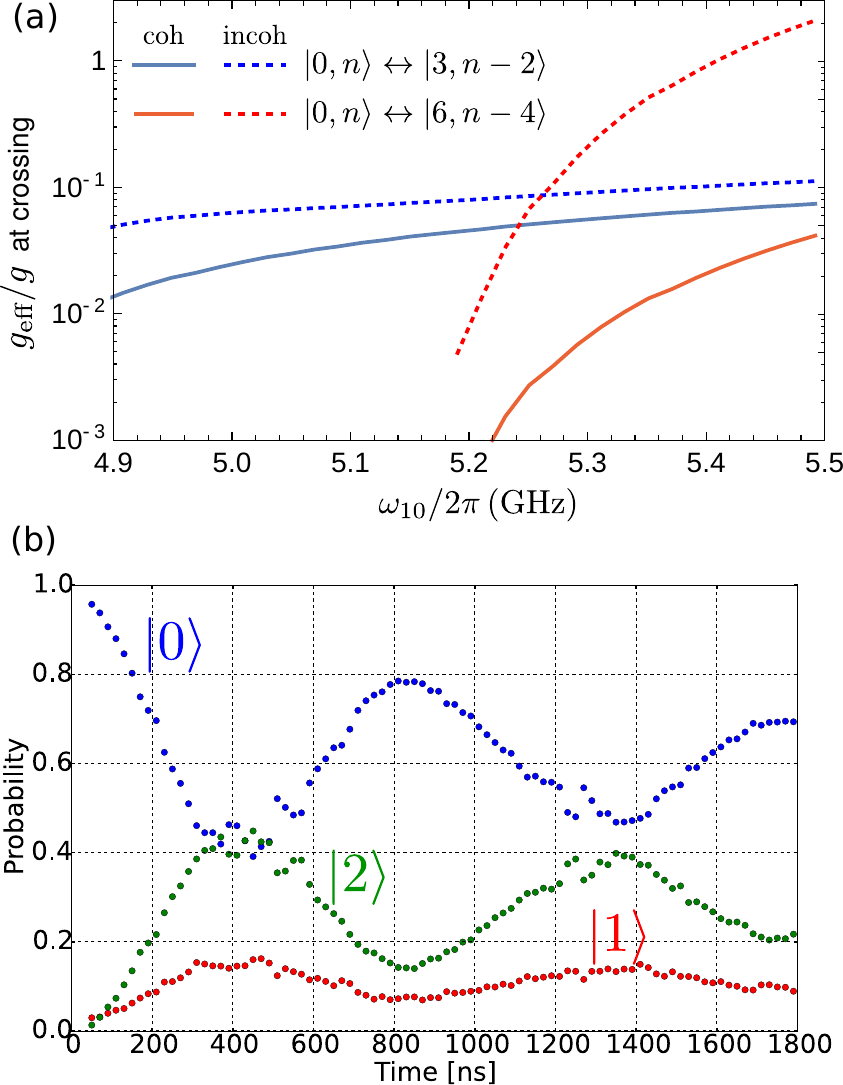}
{
(a) Effective coupling between crossing levels for different qubit frequencies.
Solid and dashed lines show coherent and incoherent effective couplings respectively.
The blue line assumes $g_{0,2}/g = 10^{-2}$.
(b) Experimental observation of Rabi oscillation between transmon levels $\ket{0}$ and $\ket{2}$.
}
{fig.g_eff}

This cancellation is probably not so efficient in the real physical system.
When the transmon is in an upper state, it is more sensitive to noise sources (such as charge noise) and therefore experiences increased dephasing.
This and the relatively low coherence of the resonator ($1/\kappa_r \approx 37\,\text{ns}$) may suppress coherence between the different paths contributing to Eq.\ \eqref{eq:g-eff-full}.
While it is difficult to accurately calculate the effective coupling $g_\text{eff}$ while accounting for decoherence, we can estimate the upper bound of the resulting $g_\text{eff}$ as the fully incoherent sum of the terms in Eq.\ \eqref{eq:g-eff-full},
\begin{align}\label{eq:g-eff-inc}
g_\text{eff}^\text{incoh} = \bigg(
& \sum \nolimits_l
\abs{ c_l^{(0,n)} \hbar g_{l,l+1} \sqrt{n-l+1} \, \left[ c_{l+1}^{(k,n-k+2)} \right]^* }^2 \nonumber \\
+ &\sum\nolimits_l
\abs{ c_l^{(0,n)} \hbar g_{l,l+3} \sqrt{n-l} \, \left[c_{l+3}^{(k,n-k+2)} \right]^* }^2 \bigg)^{1/2} \, .
\end{align}
The red dashed line in Fig.\,\ref{fig.g_eff}\,(a) shows $g_\text{eff}^\text{incoh}$ for the $\overline{\ket{0,n}} \rightarrow \overline{\ket{6,n-4}}$ transition.
We expect that the effective couplings in real system are between the results for fully coherent and fully incoherent cases (solid and dashed lines).
The experimental feature B (which corresponds to the transition $\overline{\ket{0,n}} \rightarrow \overline{\ket{6,n-4}}$) can be well explained by effective coupling on the order of 1 MHz, which is in agreement with these theoretical values (note that $g/2\pi \approx 87\,\text{MHz}$).

As discussed in the main text, the experimental feature A can be explained only if the state can transition between neighboring RWA strips (differing in total excitation number by 1).
However, if the transmon potential were exactly left/right symmetric, as is usually assumed, then $g_{k, k+2}=0$, and this transition is forbidden.
Therefore, to explain the feature A, we must assume that the transmon's symmetry is broken, leading to the additional non-RWA terms given in Eq.\ \eqref{eq:H_nonRWA-2}.
We calculated the effective coupling at the $\overline{\ket{0,n}} \rightarrow \overline{\ket{3,n-2}}$ resonance, hypothesizing that $g_{k,k+2} = 0.01 \, g \, \sqrt{(k+1)(k+2)}$ (i.e., $1\%$ violation of the selection rule).
The coupling for a coherent process is calculated via Eq.\ \eqref{eq:g-eff-c}, which for the transitions $\overline{\ket{0,n}} \rightarrow \overline{\ket{k,n-k+1}}$ between the nearest-neighbor RWA strips produces
\begin{equation}\label{eq:g-eff-full-2}
g_\text{eff}^\text{coh} = \sum\nolimits_l c_l^{(0,n)} \hbar g_{l,l+2} \sqrt{n-l} \, \left[ c_{l+2}^{(k,n-k+1)} \right]^* \, .
\end{equation}
The numerical result, indicated by the solid blue line in Fig.\,\ref{fig.g_eff}\,(a), shows that this $1\%$ violation of the selection rule yields an effective coupling of a few MHz, which is large enough to explain the experimental feature A.
The coupling becomes a few times larger if we assume the fully incoherent sum of the contributions from the paths in Eq.\ (\ref{eq:g-eff-full-2}) (constructed similarly as Eq.\ (\ref{eq:g-eff-inc}))-- see the dashed blue line in Fig.\,\ref{fig.g_eff}\,(a).
However, since the qubit state $\ket{3}$ is not supposed to experience a significant level of decoherence, we believe that the solid blue line is more relevant to the experimental situation than the dashed blue line.
It is interesting to note that the difference between the dashed and solid blue lines is much smaller than between the dashed and solid red lines, indicating that the cancellation of terms in Eq.\ (\ref{eq:g-eff-full-2}) is not as efficient as in Eq.\ (\ref{eq:g-eff-full}).
This is because for the transition $\overline{\ket{0,n}} \rightarrow \overline{\ket{3,n-2}}$ there are only two main terms in Eq.\ (\ref{eq:g-eff-full-2}): those involving $g_{0,2}$ and $g_{1,3}$. 

We experimentally looked for and actually observed the selection rule violation for $g_{0,2}$ by directly driving Rabi oscillations between transmon levels $\ket{0}$ and $\ket{2}$, as shown in Fig.\,\ref{fig.g_eff}\,(b).
By comparing the $\ket{0} \rightarrow \ket{2}$ Rabi oscillation period against the $\ket{0} \leftrightarrow \ket{1}$ Rabi oscillation period, and correcting for the differing microwave amplitude needed to drive those two transitions, we found experimentally that $g_{0,2} / g \simeq 10^{-2}$, surprisingly in good agreement with the guessed value.
We emphasize that the experimental value of $10^{-2}$ should be considered only as an order of magnitude estimate.

We can offer only speculations about the possible physical mechanism behind the broken symmetry in the transmon.
For example, it could result from SQUID asymmetry under external flux \cite{Dumur2016} or from a gradient of the magnetic field which couples to oscillating current in the circuit.
However, these mechanisms are not investigated here and will be the subject of further studies.

\section{TLS-assisted transitions}

It is well known that microscopic defects in the materials comprising the transmon circuit can act as two level systems (TLS) and lead to qubit relaxation \cite{Martinis2005}.
This relaxation can depend on the number of photons $n$ in the resonator because of the ac Stark shift.
Since ac Stark shift is approximately $\delta \omega_{10} = -2 \abs{\chi} n \simeq - (\abs{\eta}/2) (n/n_c)$, the change of the qubit frequency is quite significant ($\sim \eta \approx -200 \, \text{MHz}$) when $n$ is comparable to $n_c$.
Therefore, even if the bare qubit frequency is chosen away from the TLS frequencies, it is possible that the qubit frequency will cross a TLS during measurement with a moderate value of $n / n_c$.
In fact, we have experimentally observed this effect by comparing the transmon relaxation rate as a function of $\omega_{10}$ with $n=0$ against that same relaxation rate during dispersive measurement. 
We found that the ac Stark shift induced by the resonator photons during dispersive measurement pushes the transmon into resonance with TLS's and therefore increases the relaxation rate (data not shown).
Of course, increased relaxation degrades the fidelity of the quantum state measurement, so these crossings should be avoided.

Interestingly, coupling between the transmon and TLS's may also lead to transitions of the transmon to \textit{higher} levels, similar to the effect of the non-RWA couplings associated with resonator.
The level crossings associated with TLS's produce features similar to those produced by the non-RWA processes, such as dependence on $\Delta$.

\quickfig{\columnwidth}
{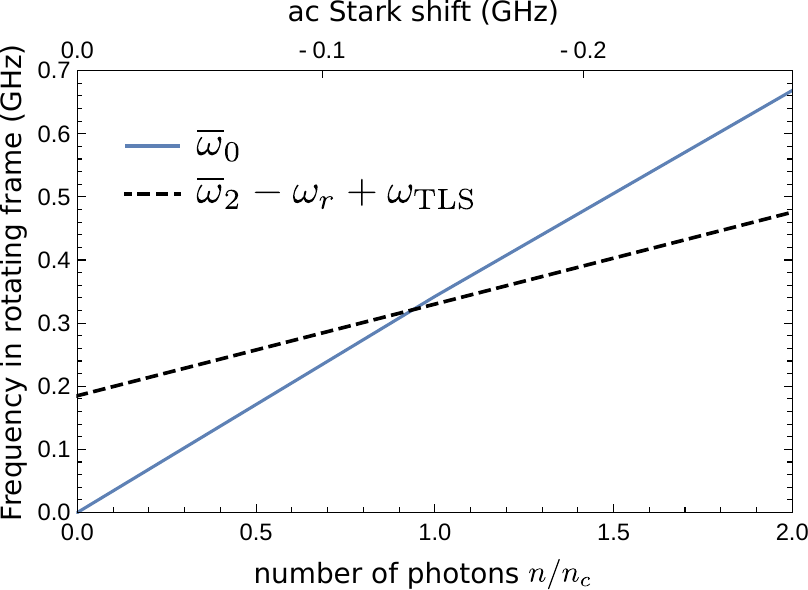}
{
Example of a resonance between transmon and a TLS.
For a TLS with frequency $10 \,\text{GHz}$, the level crossing occurs between $\overline{\ket{0,n}}\ket{0}_\text{TLS}$ and $\overline{\ket{2,n-3}}\ket{1}_\text{TLS}$.
}
{fig.TLS}

For example, the transmon can be excited from $\ket{0}$ to $\ket{2}$ via the following virtual process: $\ket{0,n}\ket{0}_\text{TLS} \rightarrow \ket{1,n-1}\ket{0}_\text{TLS} \rightarrow \ket{2,n-2}\ket{0}_\text{TLS} \rightarrow \ket{3,n-3}\ket{0}_\text{TLS} \rightarrow \ket{2,n-3}\ket{1}_\text{TLS}$.
This process requires $\omega_\text{TLS} \approx \omega_r +2 \abs{\Delta} +\abs{\eta}$ (the exact value is a little larger because of the level repulsion -- see Fig.\,\ref{fig.TLS}).
The effective coupling for these resonances can be large enough to yield noticeable population transfer at lower photon numbers than for the non-RWA resonances.
The example shown in Fig.\,\ref{fig.TLS} has a TLS with a frequency of $10\,\text{GHz}$ and the resonance for the process described above occurs at $n/n_c \approx 1$.
This value is sufficient for a noticeable amplitude of the bare state $\ket{3,n-3}$ ($c_3^{(0,n)} \approx 0.03$) and therefore a noticeable effective coupling for the process.

A TLS-assisted qubit transition from  $\ket{0}$ to $\ket{1}$ requires only population of the bare state $\ket{2,n-2}$, and therefore the effective coupling becomes significant at values of $n/n_c$ smaller than for the transition $\ket{0} \rightarrow \ket{2}$.
For example, for the parameters, corresponding to the peak in the $\ket{1}$ probability (red line) in Fig.\,2\,(c) of the main text ($n/n_c\approx 1.7$), the amplitude of the $\ket{2}$ component is quite significant, $c_2^{(0,n)} \approx 0.2$.
Therefore, even a weak coupling between the transmon and a TLS with frequency $\omega_\text{TLS}/ 2\pi \approx 8.4\,\text{GHz}$ can explain this experimental peak.
Note that when the TLS is sufficiently incoherent (e.g., because of fast energy relaxation), then the resonance condition could transform into a threshold-like condition, i.e., it should be enough energy to excite the TLS, also exciting the qubit, by transferring two photons from the resonator into the qubit-TLS system. 

With increasing $n/n_c$ and therefore increasing population of bare states $\ket{k,n-k}$, the number of possible TLS-assisted processes becomes larger (involving more final states), which increases the possibility of a transition away from the initial qubit state.
We guess that the TLS-assisted processes may be responsible for the usual deterioration of qubit measurement fidelity in many experiments when increasing $n$ becomes comparable to $n_c$ (causing either excitation or relaxation of the transmon state).


\begin{thebibliography}{34}
\bibitem [{qub()}]{qubit_t1_note}%
  \BibitemOpen
  \href@noop {} {}\bibinfo {note} {The transmon $T_1$ is between
  $20\,\mu\text{s}$ and $40\,\mu\text{s}$ for the $\omega_{10}$ values used, so
  the resonator ring-down incurs little qubit decay.}\BibitemShut {Stop}%
\end{thebibliography}
\end{document}